\documentclass[conference]{IEEEtran}
\IEEEoverridecommandlockouts
\usepackage{cite}
\usepackage{lineno,hyperref}
\usepackage{amsmath,amssymb,amsfonts}
\usepackage{algorithmic}
\usepackage{graphicx}
\usepackage{textcomp}
\usepackage{xcolor}
\usepackage{threeparttable}
\usepackage{booktabs}
\usepackage{multirow}
\usepackage{tikz}
\usepackage{colortbl}
\usepackage{footmisc}
\usepackage{verbatimbox}
\usepackage{balance}
\usepackage{flushend}
\pdfoutput=1

\newcommand*\circled[1]{\tikz[baseline=(char.base)]{
            \node[shape=circle,fill,inner sep=1.2pt] (char) {\textcolor{white}{#1}};}}
\newcommand{\project}[1]{{\texttt{\small{{#1}}}}}
\newcommand{\rqn}[1]{\texorpdfstring{RQ\textsubscript{#1}}{RQ#1}}
\newcommand{\finding}[1]{\textbf{{#1}}}
\newcommand{\urlrp}{\url{https://github.com/sftdsoftware/icsme}}

\def\BibTeX{{\rm B\kern-.05em{\sc i\kern-.025em b}\kern-.08em
    T\kern-.1667em\lower.7ex\hbox{E}\kern-.125emX}}
    
\begin{document}

\title{
Do practitioners intentionally self-fix \\ Technical Debt and why?
}

\author{Jie Tan, Daniel Feitosa, Paris Avgeriou \\
\IEEEauthorblockA{
University of Groningen, The Netherlands \\
\{j.tan, d.feitosa, p.avgeriou\}@rug.nl}
}

\maketitle

\begin{abstract}

The impact of Technical Debt (TD) on software maintenance and evolution is of great concern, but recent evidence shows that a considerable amount of TD is fixed by the same developers who introduced it; this is termed \emph{self-fixed TD}. This characteristic of TD management can potentially impact team dynamics and practices in managing TD. However, the initial evidence is based on low-level source code analysis; this casts some doubt whether practitioners repay their own debt intentionally and under what circumstances. To address this gap, we conducted an online survey on 17 well-known Java and Python open-source software communities to investigate practitioners' intent and rationale for self-fixing technical debt. We also investigate the relationship between human-related factors (e.g., experience) and self-fixing. The results, derived from the responses of 181 participants, show that a majority addresses their own debt consciously and often. Moreover, those with a higher level of involvement (e.g., more experience in the project and number of contributions) tend to be more concerned about self-fixing TD. We also learned that the sense of responsibility is a common self-fixing driver and that decisions to fix TD are not superficial but consider balancing costs and benefits, among other factors. The findings in this paper can lead to improving TD prevention and management strategies. 

\end{abstract}


\section{Introduction}

The term \emph{technical debt} (TD) is commonly used to express the adoption of suboptimal solutions in software development, that impact maintainability and evolvability  ~\cite{Li2015,kruchten_2012}.
To help practitioners in managing TD, researchers have investigated potential causes and effects of TD~\cite{rios_2020ESE_causes,behutiye_2017IST,besker_2018ICSME}, its various types (e.g., Code~\cite{palomba_2018ESE,di_2018SANER_Code}, Design~\cite{abidi_2021TOSEM,sharma_2020ESE}, and Test Debt~\cite{spadini_2018ICSME_test,de_2019MSR}) and the role of tools~\cite{avgeriou_2020,lenarduzzi_2021JSS}.
The community has also invested effort into studying specific forms of TD such as \emph{self-admitted TD}, i.e., when developers acknowledge introducing TD themselves (e.g., in source code comments)~\cite{Maldonado_ICSME2017_SATD,iammarino_2019ICSME}. More recently, there is work on \emph{self-fixed TD}, i.e., when developers repay TD that they introduced themselves~\cite{Tan2020}.

Although investigating self-fixed TD is still rather preliminary, initial findings are noteworthy. For example, Tan et al.~\cite{Tan2020} found that more than two-thirds of TD items in source code are self-fixed, and that Defect Debt tends to be self-fixed frequently, while Test Debt and Design Debt are likely to be repaid by other developers~\cite{Tan2020}.
However, current research on this topic, similarly to most work on TD, is based on source code analysis and not on the opinion of software developers. Consequently, there is no evidence of whether developers consider the differences between various debt types when repaying their own debt and what else they take into account.
We advocate that, in order to comprehend the phenomenon of self-fixed TD more thoroughly, it is paramount to consider the practitioners' intent and experiences. 

To address this shortcoming, this paper reports on a study aiming at examining the phenomenon of self-fixed TD from the perspective of practitioners. To this end, we conducted a survey with 17 well-known Java and Python open-source software communities, receiving 181 valid responses (89 for Java and 92 for Python). The survey included questions on the practitioners' attitude towards introducing and repaying TD in general and regarding five types of debt: Code Debt, Defect Debt, Design Debt, Documentation Debt and Test Debt. These five types have been extensively investigated in other studies~\cite{baldassarre_2020IST,Digkas2018,SATDMSR2016}, and were also the focus of a related work on self-fixed TD~\cite{Tan2020}. The survey also looks into practitioners' reasons to introduce and self-fix TD, as well as human factors that may influence their responses (e.g., experience and contribution levels).

This study offers a number of contributions. First, it provides insights on the relevance of self-fixed TD to practitioners and the recurrence with which they address their own debt. Second, it elaborates on the circumstances that lead to self-fixing, also providing a list of reasons (both technical and non-technical) and highlighting how these associate with reasons to introduce TD. 
These contributions can benefit both researchers and practitioners. On the one hand, researchers can build on this evidence to fine-tune TD-related tools and further study the phenomenon of self-fixed TD. On the other hand, practitioners can gain a broader understanding of the intent for introducing and self-fixing TD, which can help them devise or refine TD prevention and management strategies.

The remaining sections are organized as follows. The paper starts by summarizing related work and contextualizing our study in Section~\ref{sec:RelatedWork}. Next, Section~\ref{sec:Research_approach} reports on the goal, research questions and methodology of the study. Then, the results are presented in Section~\ref{sec:Results}, while Section~\ref{sec:Discussion} elaborates on their interpretation and implications to researchers and practitioners. Section~\ref{sec:ThreatsToValidity} reports on the threats to the validity of our study, leading to Section~\ref{sec:Conclusion} where we conclude the paper and outline directions for future work.


\section{Related Work}
\label{sec:RelatedWork}

Recent studies have shown that a large percentage of TD is paid back during software evolution~\cite{Tan_JSME2020,Digkas2018,lenarduzzi_2019TechDebt,baldassarre_2020IST}. Tan et al.~\cite{Tan2020}, 
established that a considerable amount of TD items are repaid by the same developers who introduced them. 
More recently, a number of studies focused on self-admitted technical debt (SATD): TD items that are introduced intentionally and explicitly documented in code comments~\cite{potdar_2014ICSME,SATDMSR2016,fucci_2020ICSME}.
Some related findings indicate that the majority of SATD is removed by the same developer who introduced it~\cite{Maldonado_ICSME2017_SATD,SATDMSR2016,liu_ESE2021}. However, all these previous studies drew conclusions only through source code analysis (including source code~\cite{baldassarre_2020IST} and code comments~\cite{liu_ESE2021}) and focused only on one language (i.e., either Python~\cite{Tan2020} or Java~\cite{Maldonado_ICSME2017_SATD}). In our study, we conducted a survey to investigate the practitioners' perspective on how they self-fix TD for both Java and Python. 

Other studies have focused on the rationale in TD management, e.g., the reasons to introduce or (not) repay TD. For example, Rios et al.~\cite{rios_2020ESE_causes} reported that non-technical reasons, e.g., \textit{deadline} and \textit{inappropriate planning}, play a significant role in the occurrence of TD items. Some studies also pointed out that the leading causes of technical debt are architectural choices~\cite{ernst_2015FSE} and that developers are frequently forced to introduce new TD due to already existing TD~\cite{besker_2018TechDebt}. In addition, Maldonado et al.~\cite{Maldonado_ICSME2017_SATD} found that developers self-admit TD to track potential future bugs, code that needs improvements or areas to implement new features, while they mostly remove SATD when they are fixing bugs or adding new features. Some SATD may also be accidentally removed when entire classes or methods are dropped~\cite{Zampetti_MSR18}. Researchers have also investigated the reasons for not paying TD back, e.g., \textit{lack of organizational interest}, \textit{low priority on the debt}, \textit{focus on short-term goals}, \textit{cost}, and \textit{lack of time}~\cite{freire_2020EASE}. In contrast to these studies, we focus on the reasons influencing developers' decisions to deal with their own debt, i.e., why they introduce and self-fix TD.

There is also work on the influence of human factors on incurring and managing TD. Some findings indicate that most developers that incur TD have low project-related experience~\cite{Amanatidis_XP2017} while developers' seniority and commit frequency are negatively correlated with the amount of introduced TD~\cite{Alfayez_TechDebt18}. In addition, Besker et al.~\cite{besker_2018ICSME} found that some organizational factors, such as developers' experience and startup founders' knowledge, can influence the intentional accumulation of TD. However, another study found that the developers' participation level (i.e., the total lines of code edited by a developer) and their experience in the project positively correlate with the amount of TD introduced by them, while communication skills have
barely any impact on TD~\cite{salamea_2019QRS}. In contrast to these studies, we focused on remediation of TD rather than its introduction, and in the context of self-fixed TD.


\section{Research approach}
\label{sec:Research_approach}

The goal of our study, described according to the Goal-Question-Metrics (GQM) approach~\cite{Solingen2002}, is to 
``\textit{analyze} practitioners' attitudes \textit{for the purpose of} investigating self-fixed TD \textit{with respect to} the intent and rationale for remediating it and the influence of human factors in doing so \textit{from the point of view of} software practitioners \textit{in the context of} open-source software development''. 
In the following subsections, we describe the research questions, the survey design, participants and projects selection, as well as data analysis in detail.

\subsection{Research questions}
\label{subsec:Research_questions}

We further refine the goal of this study into the following research questions:

\textbf{\rqn{1}: Do practitioners intentionally self-fix TD?}

This research question aims at investigating to what extent developers repay their own debt intentionally. More specifically, we analyze practitioners' opinions to examine which types of debt they self-fix intentionally.
The results can shed further light on how developers prioritize the remediation of technical debt. For example, developers may introduce specific types of debt with the intention of repaying it soon thereafter, thus prioritizing this item above others.

\textbf{\rqn{2}: What reasons motivate practitioners to introduce and self-fix TD?}

This research question aims at identifying practitioners' rationale for introducing and subsequently repaying their own debt. 
Analyzing the possible reasons can help to manage debt (in the case that TD is unavoidable) or prevent it (if possible). An example of management is to build tools that account for and highlight the conditions under which developers are more likely to repay a debt item. An example of prevention is to identify those reasons for introducing debt, that can be eliminated depending on a particular team's culture and composition, e.g., by making a more fine-grained planning or avoiding overloading the team.

\textbf{\rqn{3}: How do human factors influence practitioners in self-fixing TD?}

Practitioners in different roles, with different experience and contribution levels may have different attitudes towards self-fixing TD. For example, developers with a leading role tend to monitor and manage the development tasks more often than others~\cite{oliveira_2020emse,trendowicz_2009}, which may reflect in a stronger attitude towards repaying their own debt. Also, practitioners with more experience are more likely to repay technical debt~\cite{SATDMSR2016}. Thus, investigating the relationship between characteristics of practitioners and how they self-fix TD can help comprehend and subsequently support their decision-making processes when managing TD. For example, team composition can be adjusted to foster TD remediation by taking into account the diversity of roles, experiences, and cultures. In addition, researchers can also use the results to offer different tool options for practitioners with different characteristics.

\subsection{Survey Design}
\label{subsec:Survey_design}

According to our goal and the guidelines for selecting empirical methods in software engineering research~\cite{easterbrook_2008}, we decided to use a survey. A survey is a comprehensive research method for describing, comparing, or explaining knowledge and behavior~\cite{fink_2003survey}; it thus suits our focus on collecting and explaining practitioners' perceptions on self-fixed TD. More specifically, we decided to use a web-based questionnaire to conduct the survey because it is time- and cost-effective, and also suitable for collecting data from a large number of developers in geographically diverse locations~\cite{Lethbridge2005}. 

\begin{figure}[htbp]
\centerline
{\includegraphics[clip, trim=2cm 22cm 5.6cm 5cm, width=0.49\textwidth]{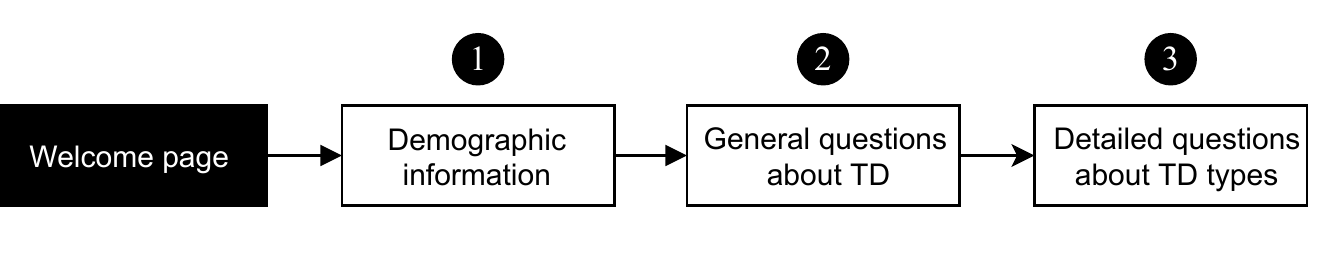}}
\caption{Design of the survey in our study.}
\label{fig:design}
\end{figure}

The flow of the survey design is presented in Fig.~\ref{fig:design}. The black box represents the introductory welcome page, and the numbered white boxes depict steps in which participants answer questions. The data collection was planned to last two months and the questionnaire was made accessible online between February and March 2021. To mitigate privacy concerns, all invited participants were informed about the survey with an invitation text describing the target open-source project and programming language (see Section \ref{subsec:Participants_Selection}), together with the motivation and goal of this study. Moreover, we did not collect any identifiable information in the survey.
The templates used for communication are available in the replication package\footnote{\urlrp}.

As shown in Fig.~\ref{fig:design}, the survey started with a welcome page explaining the structure of the questionnaire, and that the survey is anonymous. On step \circled{\footnotesize{1}}, the survey starts with a form with demographic questions. In particular, we inquired about the participants' role in the target software project (e.g., developer, tester), their educational background, the number of years they have been working in software development, the number of years they have been working on the project, and the number of commits they have contributed to the project. Moreover, since this survey targeted a broad audience, we wanted to assess how balanced the population is and, thus, we also asked for the participants' sex and country (as optional questions).

Subsequently, the survey included five general questions (see Table~\ref{tab:TD_general_question}) related to technical debt in step \circled{\footnotesize{2}}. But before asking any questions, we start by describing three terms to set a baseline for common understanding: what \textit{technical debt} is, and what it means to \textit{introduce} it and \textit{repay} it. We then ask GQ1 to identify whether the participants are familiar with the aforementioned descriptions of these terms.
Next, 
to investigate whether practitioners are aware of introducing and repaying their TD, we record the frequency (from \textit{always} to \textit{never}), with which practitioners introduced (GQ2) and self-fixed (GQ4) TD. We also ask participants the reasons why they have introduced (GQ3) and self-fixed (GQ5) TD. The options for GQ3 stem from the results of a previous study~\cite{ESEM18_Rios}, which reported the top ten most cited TD causes as informed by 107 participants. GQ5 is an open-ended question as there was no established knowledge in the literature on reasons for self-fixing TD. Finally, for GQ3, in addition to the listed options, we add the option \textit{Other} with a text field for participants to provide more details.

\begin{table}[htbp]
  \centering
  \caption{TD general questions of the survey}
  \setlength{\tabcolsep}{3pt}
    \addvbuffer[-3pt -10pt]{
    \begin{tabular}{cp{7.8cm}}
    \toprule
    \textbf{No.} & \textbf{TD General Question Description}\\
    \midrule
    GQ1 & Were you familiar with the presented description of technical debt before this questionnaire?\\
    \midrule
    GQ2 & How often do you introduce technical debt? \\
    \midrule
    GQ3 &For what reasons have you introduced technical debt? \\
    \midrule 
    GQ4 & How often do you repay your own technical debt? \\
    \midrule
    GQ5 & For what reasons do you repay your own technical debt? \\
    \bottomrule
    \end{tabular}
    }
    \label{tab:TD_general_question}
\end{table}

The questions in step \circled{\footnotesize{3}} covers all five types of debt. For each debt type, we provide the definition and ask the participants about their recollection of 
(DQ1) \textit{noticing} and 
(DQ2) \textit{self-fixing}
TD items. For that, we use a Likert scale with the options \textit{definitely not}, \textit{probably not}, \textit{I am not sure}, \textit{probably yes}, and \textit{definitely yes}.

\subsection{Projects and Participants Selection}
\label{subsec:Participants_Selection}

For this study, we selected projects that are written in either Java or Python, which are ranked as the top two most popular object-oriented programming languages\footnote{According to the Tiobe Index, which is one of the best-known indexes of programming languages popularity: \url{https://www.tiobe.com/tiobe-index/}, visited March 2021}. Although both of them are strongly typed languages, Java is statically typed while Python is dynamically typed. Because of the dynamic features, Python projects are more change-prone, resulting in developers spending more effort on software maintenance~\cite{PythonFeatureSEKE} and software quality improvement~\cite{PythonQuality}. Thus, selecting systems written in these two programming languages, helps to increase the external validity of our study. Moreover, 
investigating both languages can help to compare the findings to the previous results that focus on source code analysis~\cite{Digkas2018, Tan2020}. In the following, we describe the process of selecting projects and participants for the survey.

We started by compiling a list of systems that have been widely used in the studies that investigated technical debt through source code analysis (see Section \ref{sec:RelatedWork}). From this list, we filtered out projects with less than 5,000 commits or 200 contributors, to ensure that a sufficient number of practitioners are familiar with the maintenance of the projects for this survey. Ultimately, the following eleven projects were selected: \project{Camel}~\cite{Maldonado_ICSME2017_SATD,iammarino_2019ICSME}, 
\project{Cassandra}~\cite{wehaibi_2016SANER}, 
\project{Chromium}~\cite{wehaibi_2016SANER,bellomo_2016MSR}, \project{Elasticsearch}~\cite{flisar_2018SEAA}, \project{Gerrit}~\cite{Maldonado_ICSME2017_SATD,iammarino_2019ICSME}, \project{Hadoop}~\cite{yan_2018TSE,zampetti_2020SANER}, 
\project{HBase}~\cite{palomba_IST2018}, 
\project{Guava}~\cite{lerina_2019TechDebt,tsoukalas_2020JSS}, \project{Jenkins}~\cite{tsoukalas_2020JSS,lerina_2019TechDebt}, \project{RxJava}~\cite{lerina_2019TechDebt,iammarino_2019ICSME}, 
\project{TensorFlow}~\cite{liu_ESE2021}. Considering that, except for \project{TensorFlow}, these are Java projects, we still needed to supplement the list with pertinent Python projects. In addition to complying with the same thresholds for the number of commits and contributions, we chose the Python projects with the most stars on GitHub: \project{Ansible}, \project{Django}, \project{Flask}, \project{Pandas}, \project{Scikit-Learn}, and \project{Superset}. The number of stars of a GitHub repository
can be seen as a proxy of its popularity~\cite{borges_2016ICSME} as more stars indicate that more users are interested in the project~\cite{tsay_2014ICSE}. 

To obtain the target sample of participants, we extracted each contributor's name, email address, and the time of each commit from GitHub. Next, for each project, we consider the contributors who had submitted at least two commits within the last three years; this helps to target contributors who still have a recollection of their work in the project and whose email address is likely to still be in use, thus improving the response rate and the response quality. To mitigate privacy concerns, we did not retain any information after sending invitations. In total, we sent out 7638 emails.

\subsection{Data Analysis}
\label{subsec:data_analysis}

\begin{table}
  \centering
  \caption{Mapping between survey questions, variables and research questions}
  \setlength{\tabcolsep}{6pt}
  \addvbuffer[-3pt -10pt]{
  \begin{tabular}{cll}
    \toprule
    \textbf{No.} & \textbf{Variables} & \textbf{RQs}\\
    \midrule
    BQ1 & demographic information, i.e., the practitioners' role, & \\
    -- & educational background, sex, country, contribution, & \rqn{3}\\
    BQ7 & experience in the project and software development & \\
    \midrule
    GQ1 & practitioner's familiarity with technical debt & \rqn{1}\\
    GQ2 & the frequency of introducing technical debt & \rqn{2}\\
    GQ3 & the reasons why developers have introduced TD & \rqn{2}\\    
    GQ4 & the frequency of self-fixing technical debt & \rqn{2,3}\\
    GQ5 & the reasons why practitioners have self-fixed TD & \rqn{2}\\ 
    \midrule
    DQ1 & the types of debt that practitioners have noticed & \rqn{1}\\
    DQ2 & the types of debt that practitioners have self-fixed & \rqn{1,3}\\ 
    \bottomrule
  \end{tabular}
  }
  \label{tab:mapping}
\end{table}
Table~\ref{tab:mapping} presents the mapping between the questions in the survey, the collected variables and research questions. The three sets of survey questions correspond to the three steps in the survey design (i.e., step \circled{\footnotesize{1}} -- \circled{\footnotesize{3}} in Fig.~\ref{fig:design}): background (demographic) questions (BQ), general questions about technical debt (GQ), and detailed questions about TD types (DQ). In the following, we describe in detail how each of the research questions makes use of the responses.

To answer \rqn{1}, we first investigate to what extent developers are familiar with the description of technical debt from the responses of GQ1, using descriptive statistics. Next, we answer to what extent developers self-fix technical debt intentionally per TD type; for that, we use the responses of DQ2 from step \circled{\footnotesize{3}} of the survey (i.e., \textit{Do you recall self-fixing technical debt} (per type)?). Since the responses to this question use a five-point Likert scale (i.e., from \textit{definitely not} to \textit{definitely yes}---see Section~\ref{subsec:Survey_design}), the results are presented in the form of stacked bar charts and comparisons are assessed using statistical tests. We also make a supplementary discussion in conjunction with the types of debt that participants have noticed in source code (DQ1), since identifying TD items may contribute to one's knowledge of TD.

To answer \rqn{2}, i.e., the reasons why practitioners introduce and self-fix TD, we first investigate the extent to which they are aware of introducing and self-fixing debt (GQ2 and GQ4, respectively). For that, we create and analyze a heat map comparing various frequencies of TD introduction to various frequencies of self-fixing it (see Section~\ref{subsec:Survey_design}) and evaluate the association between them using several statistical tests.
Then, we qualitatively discuss the reasons provided by participants to explain why they decide to introduce (GQ3) and self-fix technical debt (GQ5). Since GQ3 provides the option \textit{Other} (with a text field) and GQ5 is an open-ended question, we followed an open-coding approach~\cite{seaman_1999TSE_qualitative,miles_1994_qualitative} to analyze the answers collected from the survey respondents; a similar approach was followed in many recent software engineering studies, e.g., ~\cite{aghajani_2020ICSE,vadlamani_2020ICSME}. To conduct open coding, the first two authors independently assigned one or more tags to each answer; each tag is meant to summarize a reason. The level of inter-rater agreement between the classifications of the two authors was measured and
conflicts were solved by a discussion with the third author. 
Since participants can provide multiple reasons to introduce and self-fix TD, the associations between reasons are presented in the form of a chord diagram and investigated using descriptive statistics.

For \rqn{3}, to explore how the human factors (characteristics of the practitioners such as roles, experience and contribution) affect their self-fixing behavior, we investigate the relationship between the demographic information of practitioners (step \circled{\footnotesize{1}} in Fig.~\ref{fig:design}) and their perspective of self-fixing TD (step \circled{\footnotesize{2}}\circled{\footnotesize{3}} in Fig.~\ref{fig:design}). For that, we first calculate, for each demographic group, the proportions of different ratings for the frequency (GQ4) and certainty (DQ2) of self-fixing TD. Then, we conducted several statistical tests to explore whether one demographic group tends to rate the frequency or certainty higher or lower than other groups.


\section{Results}
\label{sec:Results}

Overall, we obtained 184 responses to our survey and the response rate is 2.41\%. Among them, 181 responses were valid: 89 for Java and 92 for Python. The three invalid responses lack specific information as, e.g., respondents choose ``Other'' and answered ``NA'' for several questions. The distribution of the population is presented in Table~\ref{tab:characteristics}. In the following, we present the results of the study organized by research question.

\begin{table}
  \centering
  \caption{Characteristics of the respondents}
  \setlength{\tabcolsep}{2pt}
  \addvbuffer[-3pt -10pt]{
    \begin{tabular}{lrr|lrr}
    \toprule
    \textbf{Characteristics} & \textbf{Java} & \textbf{Python} & \textbf{Characteristics} & \textbf{Java} & \textbf{Python} \\
    \midrule
    \multicolumn{3}{l|}{\textbf{Main role in the project}} 
    & \multicolumn{3}{l}{\textbf{Experience in the project}} \\
    Developer* & 76.40\% & 69.56\% & \textless \ 1 year & 8.99\% & 33.70\% \\
    Software Architect & 8.99\% & 14.13\% & 1--2 years & 26.97\% & 36.96\%\\
    Team Leader & 6.74\% & 4.35\% & 3--5 years & 32.58\% & 15.22\% \\ 
    Product Owner & 3.37\% & 5.43\% & \textgreater \ 5 years & 31.46\% & 14.13\%\\  
    Tester** & 1.12\% & 2.17\% & \\
    System Administrator & 1.12\% & 1.09\% & \multicolumn{3}{l}{\textbf{Contribution to the project}} \\
    DevOps Engineer & 1.12\% & 1.09\% & \textless \ 10 commits & 11.24\% & 43.48\%\\
    Engineering Manager & 1.12\% & & 10--50 commits & 12.36\% & 14.13\% \\
    Project Manager & & 2.17\% & 50--100 commits & 10.11\% & 14.13\% \\
    & & & 100--200 commits & 12.36\% & 6.52\%\\
    \multicolumn{3}{l|}{\textbf{Educational background}} &  200--500 commits & 14.61\% & 7.61\%\\
    Master degree & 48.31\% & 30.43\% & \textgreater \ 500 commits & 39.33\% & 14.13\% \\
    Bachelor degree & 39.33\% & 36.96\% \\
    Ph.D. degree & 8.99\% & 20.65\% & \multicolumn{3}{l}{\textbf{Software development experience}} \\
    High school & 1.12\% & 4.35\% & \textless \ 5 year & 12.36\% & 21.74\% \\    
    Other / not shared & 2.25\% & 7.61\% & 5--10 years & 16.85\% & 26.09\% \\
    & & & 10--20 years & 44.94\% & 28.26\% \\    
    \multicolumn{3}{l|}{\textbf{Geographical location}} & 20--35 years & 23.60\% & 18.48\% \\
    Europe & 52.81\% & 29.35\% & \textgreater \ 35 years & 1.12\% & 4.35\% \\ 
    North America & 21.35\% & 39.13\% & Not shared & 1.12\% & 1.09\% \\
    Asia & 10.11\% & 23.91\% & \\
    South America & 5.62\% & 2.17\% & \multicolumn{3}{l}{\textbf{Respondent's sex}}\\    
    Oceania & 2.25\% & 1.09\% & Male & 87.64\% & 90.22\%\\
    Not shared & 7.87\% & 4.35\% & Female & 3.37\% & 6.52\%\\
    & & & Other & 1.12\% & \\
    & & & Not answered & 7.87\% & 3.26\%\\
    \bottomrule
    \multicolumn{3}{l}{*Developer, incl. Software Engineer} & \multicolumn{3}{l}{**Tester, incl. Quality Assurance}\\
    \end{tabular}%
    }
    \label{tab:characteristics}%
\end{table}%


\subsection{Do practitioners intentionally self-fix technical debt? (\rqn{1})}

\begin{figure}[htbp]
\centerline
{\includegraphics[width=0.4\textwidth]{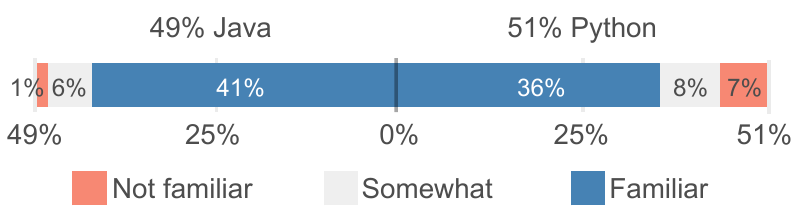}}
\caption{Respondents' knowledge level of technical debt}
\label{fig:rq1_familiar}
\end{figure}

To answer this research question, we first investigated to what extent developers are familiar with the concept of technical debt (GQ1 in Table~\ref{tab:TD_general_question}). Fig.~\ref{fig:rq1_familiar} depicts the answers through a diverging stacked bar chart to visualize the percentages that describe the distribution of responses. The left side shows the answers of the 89 Java participants (49\%), while the answers of the 92 Python participants (51\%) are presented on the right. 

According to Fig.~\ref{fig:rq1_familiar}, the vast majority (i.e., 92\%) of the participants were familiar or somewhat familiar with technical debt before this survey. This increases our confidence that the data we collect is (mostly) derived from knowledgeable individuals (a thorough discussion of threats to validity is found in Section \ref{sec:ThreatsToValidity}). In addition, Java participants seem slightly more familiar than Python participants. To further evaluate the significance of such difference, we calculated the Wilcoxon Rank Sum test~\cite{Wilcoxon_2014} ($p$-value = 0.02) and Cliff's Delta Effect Size~\cite{cliff_2014} ($delta$ = 0.1468) on the ratings of the Java and Python respondents, since we cannot assume the populations are normally distributed. The results reveal that the difference between the familiarity of Java and Python participants with TD is significant (i.e., $p$-value \textless 0.05) but the effect size is negligible (i.e., $|delta|$ \textless 0.147). Thus, different levels of familiarity with the concept of TD between Java and Python participants have almost no effect on the disparity between the populations in subsequent findings.

\begin{figure}[htbp]
\centerline
{\includegraphics[width=0.5\textwidth]{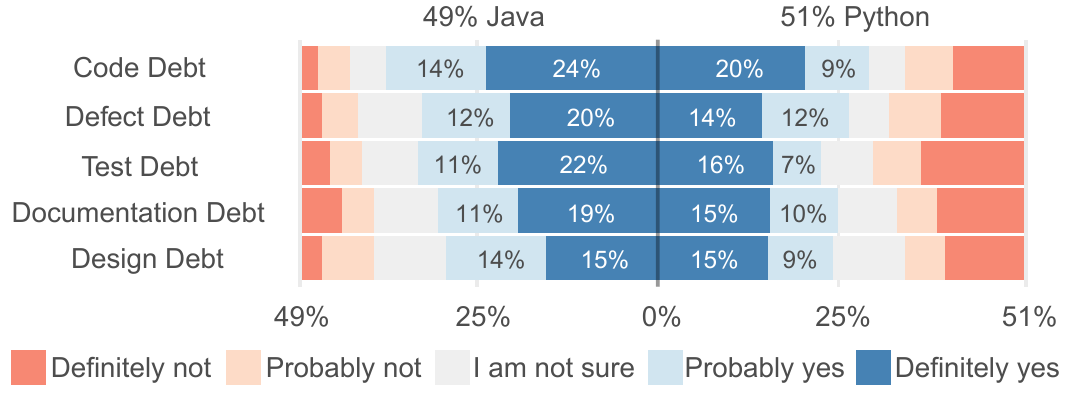}}
\caption{Distribution of self-fixed TD types}
\label{fig:rq1_survey1}
\end{figure}

Delving into the essence of the research question, we looked at whether participants recalled self-fixing each type of debt (DQ2). While incurring TD can be both intentional and unintentional, self-fixing in this case is intentional: if practitioners recall paying back their own TD items, it is safe to assume they did it on purpose (at least for those items). The answers are summarized in Fig.~\ref{fig:rq1_survey1}. This figure shows that, for each type of TD, more than half of the participants mentioned that they \textit{definitely} or \textit{probably} have self-fixed items. To investigate the significance of this observation, we first combined the answers into three categories: (1) positive (\textit{definitely} or \textit{probably} self-fixed TD); (2) neutral (\textit{I am not sure}); and (3) negative (\textit{definitely not} or \textit{probably not} self-fixed TD). Next, we conducted a Scott-Knott Effect Size Difference (ESD) test~\cite{tantithamthavorn_2016TSE_ESD} to group the three categories of answers into statistically distinct ranks based on the combined  proportions of all debt types.
As a variant of the Scott-Knott test~\cite{scott_1974}, the Scott-Knott ESD test evaluates a non-normally distributed dataset and merges any two statistically distinct groups that have a negligible effect size into a single group~\cite{tantithamthavorn_2016TSE_ESD}. The result shows that the three categories belong to different groups and, thus, the percentage differences observed are significant. Since the percentage of positive answers (category \#1) is considerably greater than the other two, it implies that~\finding{developers often display a positive attitude towards self-fixing TD}.

Looking into individual deb types, participants seem to be slightly more concerned about self-fixing Code Debt issues, less about Test Debt, and mostly uncertain about Design Debt.
To assess the significance of these differences, we conducted a Scott-Knott ESD test to group TD types into statistically distinct ranks based on the Likert scale. The result shows that with the exception of Code Debt, there is no significant distinction between debt types. Consequently, \finding{as Code Debt displays the highest positive score, it seems to be the main concern of the participants}.

Comparing the languages, we observed from Fig.~\ref{fig:rq1_survey1} that Java participants seem to be more likely to acknowledge that they have \textit{definitely} or \textit{probably} self-fixed TD, as more than half of them mentioned so for each debt type. To assess this particular observation, we conducted two Scott-Knott ESD tests (for Java and Python populations respectively) to group TD types into statistically distinct ranks based on their Likert scores. The result shows that there are three types of TD (i.e., Code Debt, Test Debt, and Defect Debt) belonging to the same group in Java, while Code Debt alone makes up one group in Python.
\finding{Although Code Debt is reaffirmed as an important concern in both languages, Test Debt and Defect Debt are as relevant as Code Debt only to the Java participants.}


\subsection{What reasons motivate practitioners to introduce and self-fix TD? (\rqn{2})}

\begin{figure}[htbp]
\centerline
{\includegraphics[width=0.5\textwidth]{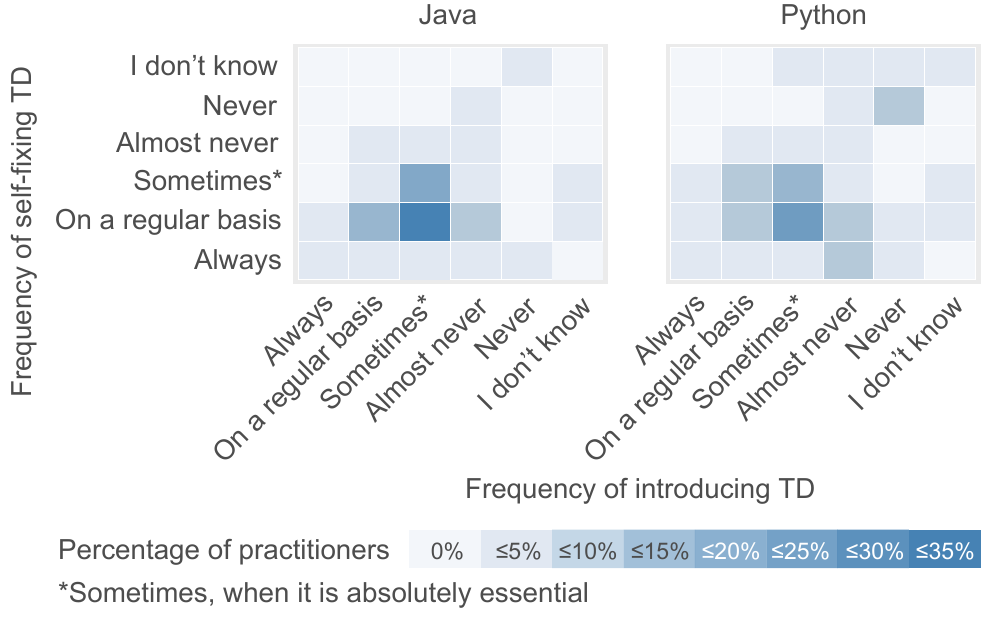}}
\caption{Percentage of participants indicating the frequency of self-fixing TD (column) for a given frequency of TD introduction (row).}
\label{fig:rq2_1}
\vspace{-3mm}
\end{figure}

To answer this research question, we first investigated the extent to which practitioners are introducing and self-fixing TD. Fig.~\ref{fig:rq2_1} shows a heat map depicting the percentages of participants who self-fixed TD at different frequencies for a given frequency of TD introduction in Java and Python, respectively. A Wilcoxon Signed Rank test~\cite{rosner_2006_wilcoxonrank} reveals no significant difference between Java and Python ($p$-value = 0.29), suggesting
that Java and Python participants introduce and self-fix TD in a similar fashion.

\begin{figure}[htbp]
\centerline
{\includegraphics[width=0.48\textwidth]{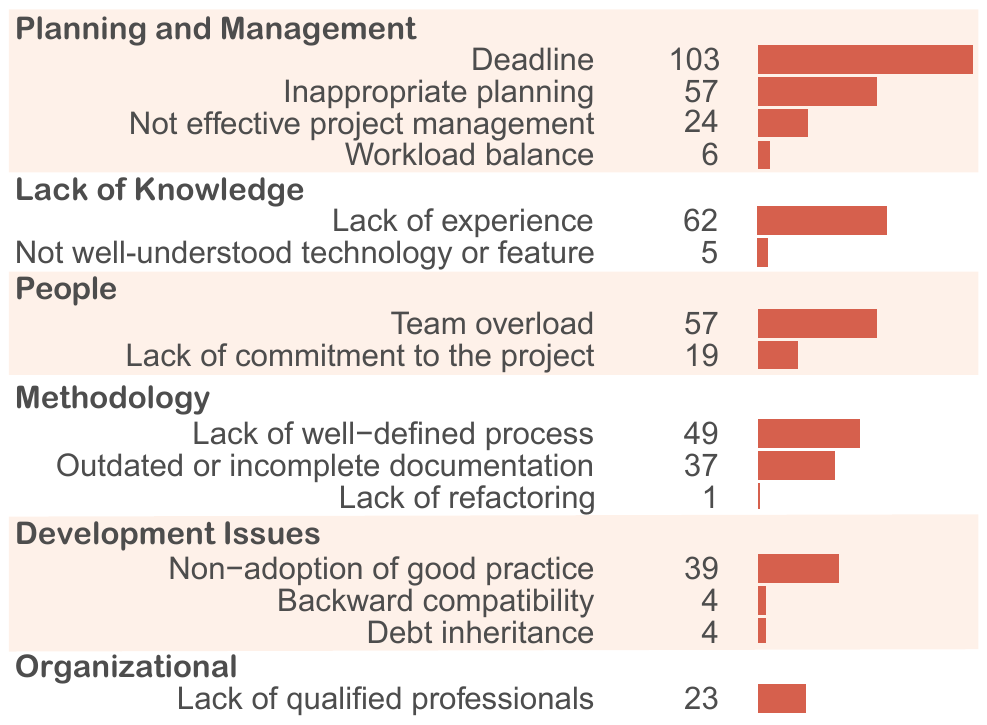}}
\caption{Reasons to introduce technical debt}
\label{fig:rq2_intro_reasons}
\end{figure}

Analyzing the population as a whole, almost half of the participants (48\%) have acknowledged that they self-fix TD \textit{on a regular basis} and a half of those (27\% of the general population) acknowledge also introducing debt sometimes only. This combination seems more recurrent than any other. To assess the significance of this observation, we conducted a Scott-Knott ESD test to group the frequencies of self-fixing into statistically distinct ranks based on introduction frequencies. The result shows participants that self-fix TD \textit{on a regular basis} belong in a distinct group, and participants that do it \textit{sometimes} in another distinct group. The results indicate that \finding{although participants sometimes have to introduce TD, when they do, they also tend to pay it back.} 

In the following, we examine the reasons that lead practitioners to introduce and self-fix TD. Regarding the former, 37 of all the 181 participants also selected the option \textit{Other} and provided custom opinions. We applied an open-coding approach and extracted one or more reasons from each custom answer that could either fit into one of the ten pre-defined reasons or become a new reason. By the end of the coding, five new reasons had been created. Fig.~\ref{fig:rq2_intro_reasons} shows the 15 reasons that can lead to TD introduction from the participants' perspective. The overall frequency of each reason is the number of participants who mentioned it. Apart from the reasons listed in the Figure, 20 participants selected that they \textit{are not aware of introducing TD}.
Another four participants indicated that \textit{TD introduction is inevitable}, e.g., ``all code is technical debt'' and ``TD introduction is a part of software development''; one of them mentioned also the reason \textit{non-adoption of good practice}.


To facilitate the interpretation of the reasons to incur TD, we grouped them into more general categories, which we adopted from Rios et al.~\cite{rios_2020ESE_causes}. The categories are also shown in Fig.~\ref{fig:rq2_intro_reasons}, from which we observe that \textit{deadline} is cited by more than half of the participants, indicating that it is a factor that normally contributes to the appearance of TD. During the open-coding, some custom answers were coded as \textit{deadline}, e.g., ``it is evaluated and decided to be worth incurring this technical debt to finish the task earlier''. 
Furthermore, \textit{lack of experience}, \textit{inappropriate planning}, and \textit{team overload} are other reasons cited by at least 30\% of the participants. Among them, \textit{lack of experience} also includes custom answers such as ``lack of learning'' and ``not fully understanding how the project will evolve over time''. 

The five new reasons that appeared during the open-coding are: \textit{workload balance} referring to the trade-off between the cost and benefit of introducing TD, e.g., ``prioritize mainline tasks and de-prioritize secondary tasks'', and ``increasing tech debt is desirable''; \textit{not well-understood technology or feature}, e.g., ``experiments to introduce new technologies'' and ``a feature might not be well understood and it is more cost effective to do it in a suboptimal way''; \textit{debt inheritance} referring to new debt introduced due to a pre-existing TD item, e.g., ``working in an area that is already burdened with technical debt''; \textit{backward compatibility}, e.g., ``backward compact and minor version release rules'' and ``backportability''; and \textit{lack of refactoring}, e.g., ``ongoing major refactoring''.

\begin{figure}[btp]
\centerline
{\includegraphics[width=0.48\textwidth]{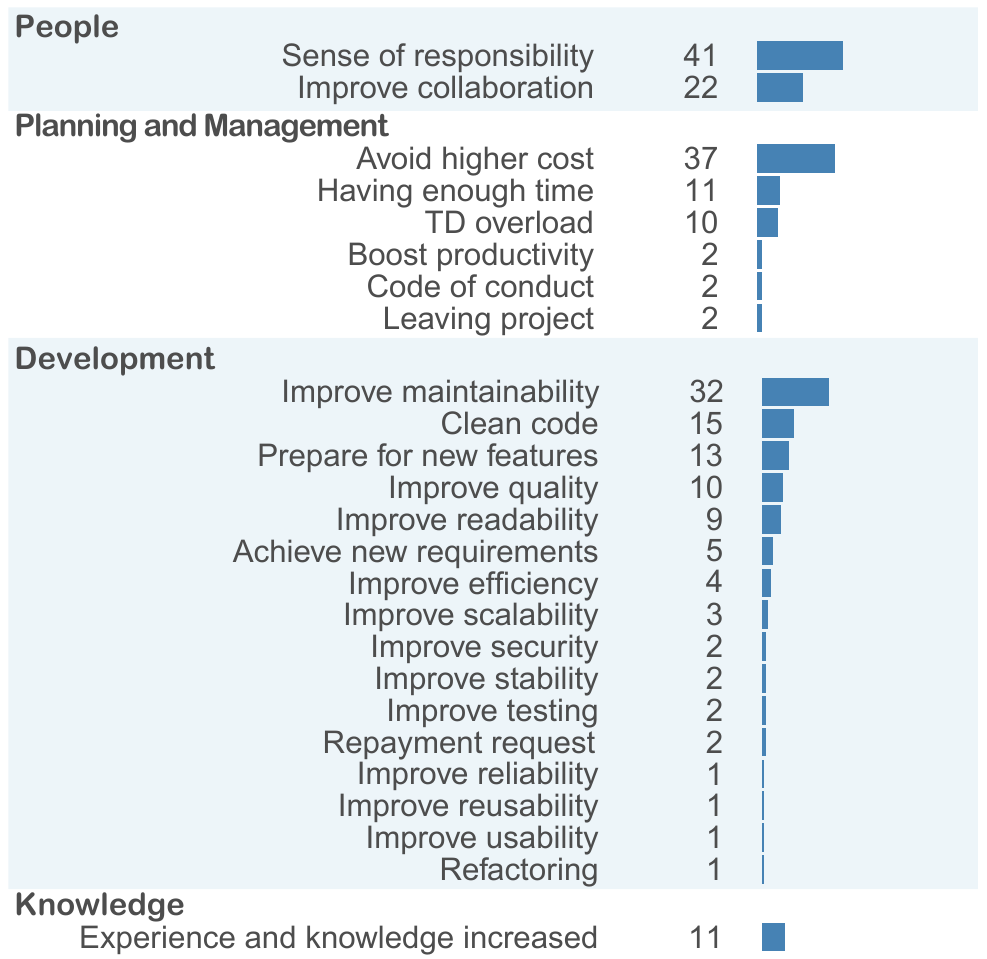}}
\caption{Reasons to self-fix technical debt}
\label{fig:rq2_self-fix_reasons}
\end{figure}

Regarding the reasons to self-fix TD, we summarized 25 reasons from the responses (all open-ended in this case), as shown in Fig.~\ref{fig:rq2_self-fix_reasons}. Similarly to Fig.~\ref{fig:rq2_intro_reasons}, we also mapped the reasons into more general categories. For that, we adopted the categories provided by Rios et al.~\cite{rios_2020ESE_causes} again, but renamed some categories to fit the context.
During the open-coding, answers were often mapped into multiple reasons. For example, one participant answered: ``improving code quality, maintainability, making new features easier to implement''. This answer was coded as \textit{improve quality}, \textit{improve maintainability}, and \textit{prepare for new features}.

After examining the reasons for self-fixing TD, we noticed that \textit{sense of responsibility} and \textit{avoid higher cost} are the most frequently mentioned, each being cited by at least 37 participants (approx. 20\%). For example, one participant mentioned ``I feel ownership over the code, also, better to get rid of it before it bites you'', which was coded as both \textit{sense of responsibility} and \textit{avoid higher cost}.

\begin{figure}[btp]
\vspace{-4mm}
\centerline
{\includegraphics[width=0.4\textwidth]{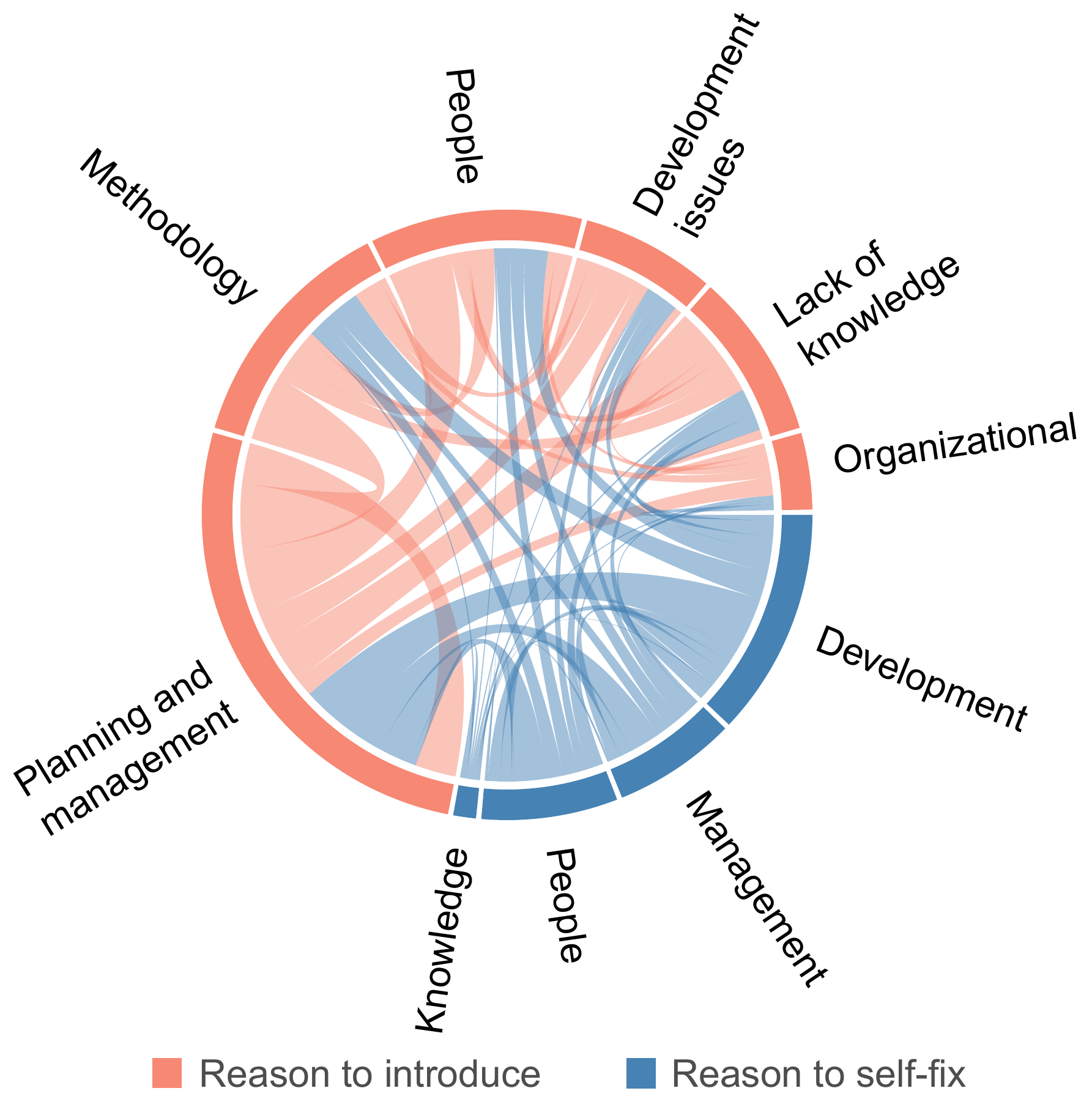}}
\caption{Association between reasons}
\label{fig:rq2}
\vspace{-2mm}
\end{figure}

Most participants mentioned multiple reasons when answering why they introduce and self-fix TD. Thus, to gain a better understanding on the relationship between these reasons,  
we calculated the number of times that every pair of categories was mentioned together (i.e., totally 1536 relations), and plotted a chord diagram~\cite{gu_2014} illustrating the relations between categories of reasons (cf. Fig.~\ref{fig:rq2}).  
The categories for introducing and self-fixing TD are colored red and blue, respectively, and the width of the links conveys the strength of the relation.

From Fig.~\ref{fig:rq2}, we can observe that among all the relations, the reasons to introduce TD (in red) seem to display the strongest relationships with themselves. In particular, the reasons related to \textit{planning and management} and \textit{methodology} were most often mentioned simultaneously (127 times), followed by \textit{planning and management} and \textit{people} (121 times).
Regarding the reasons for self-fixing TD, it is noticeable that the links between the self-fixing and introduction categories are wider than those between self-fixing reasons themselves.
In other words, the reasons for self-fixing TD were mentioned more often together with those to introduce TD than with themselves (629 vs. 112 times). Especially the reasons to self-fix TD related to \textit{development} were often mentioned when the reasons to introduce TD concerned \textit{planning and management} (108 times).

Altogether, we observe both proactive reasons (e.g., \textit{sense of responsibility}, \textit{avoid higher cost}) and reactive reasons (e.g., \textit{TD overload}, \textit{Repayment request}) to self-fix TD. More importantly, the evidence suggests that \finding{regardless of the reason that lead a practitioner to introduce TD, she is keen to repay the debt proactively} not only due to an internal driver (e.g., \textit{sense of responsibility} or \textit{improve collaboration}) but also for strategic reasons (e.g., \textit{avoid higher debt} or \textit{improve maintainability}). Further evidence that repayment is thoughtfully considered, is that most reasons to self-fix TD co-occur with reasons to introduce TD related to \textit{planning and management}.


\subsection{How do human factors influence practitioners in self-fixing TD? (\rqn{3})}

\newcommand{\dg}[1]{\texttt{#1}}

To answer this research question, we followed a process similar to that used by Wan et al.~\cite{wan_2018TSE} and Zou et al.~\cite{zou_2018TSE} to analyze 11 demographic groups derived from the \emph{team roles}, \emph{development experience}, \emph{project experience} and \emph{contribution to project}. For comparison reasons, we also consider an additional group (\dg{All}) including all participants. We explain the demographic groups in the following paragraphs.

Regarding the \emph{team roles}, one group corresponds to developers (\dg{Dev}), including software engineers. Another group consists of respondents who are involved in the project or team management and support (\dg{PM}), i.e., software architect, team leader, product owner, system administrator, DevOps engineer, engineering manager and project manager. In addition to these two groups, we also received three responses from testers. However, we did not consider testers as a separate group because the sample size does not fit the threshold for statistical analysis~\cite{siegel_1956nonparametric}, which would also increase the randomness of the results~\cite{lenth_2001some}.

Regarding the levels of \emph{software development experience}, we distinguish between respondents with: (a) high experience (\dg{ExpSdHigh}), as the 25\% with the most experience in years~\cite{zou_2018TSE}, i.e., $\geq$ 18 years in this survey; low experience (\dg{ExpSdLow}), as the 25\% with the least experience in years, i.e., $\leq$ 5 years; and medium experience (\dg{ExpSdMed}), i.e., the remaining respondents.
For \emph{project experience}, since the survey offered discrete options, we selected the maximum and minimum interval that approximates the top and bottom 25\%. In particular, we defined a high experience group (\dg{ExpProHigh}) with those that have five or more years in a project (approx. 23\%), a low experience group (\dg{ExpProLow}) with those that joined a project less than one year ago (approx. 22\%), and a medium experience group (\dg{ExpProMed}) with the remaining respondents.
The \emph{contributions} to projects are grouped in a similar fashion, as follows: high contribution level (\dg{ContribHigh}), with participants that submitted 500 commits or more (approx. 27\%); low contribution level (\dg{ContribLow}), with those that submitted up to 10 commits (approx. 28\%); and medium contribution level (\dg{ContribMed}), with the remaining respondents.

\begin{figure}[htbp]
\centerline
{\includegraphics[width=0.45\textwidth]{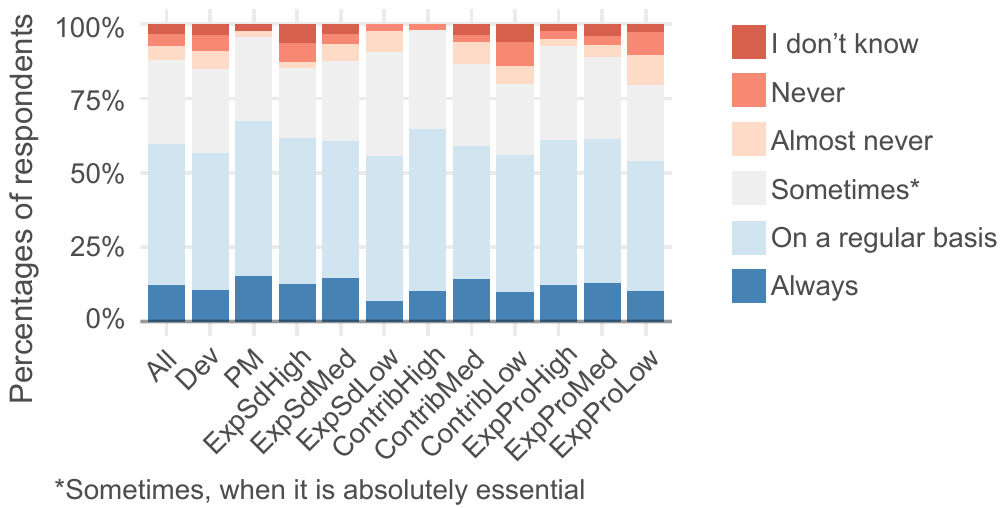}}
\caption{The percentages of how often participants in each demographic group have self-fixed TD}
\label{fig:rq3_frequency}
\end{figure}

Fig.~\ref{fig:rq3_frequency} presents the frequencies of self-fixing TD as perceived by respondents in the eleven demographic groups. As shown in the figure, the majority of participants in all of the 12 groups (79\%--97\%) mentioned they repay their TD \textit{always}, \textit{on a regular basis}, or \textit{sometimes} (\textit{when it is absolutely essential}). In particular, about half of the participants in the various groups (44\%--54\%) considered they self-fix TD \textit{on a regular basis}.

\begin{figure}[htbp]
\centerline
{\includegraphics[width=0.45\textwidth]{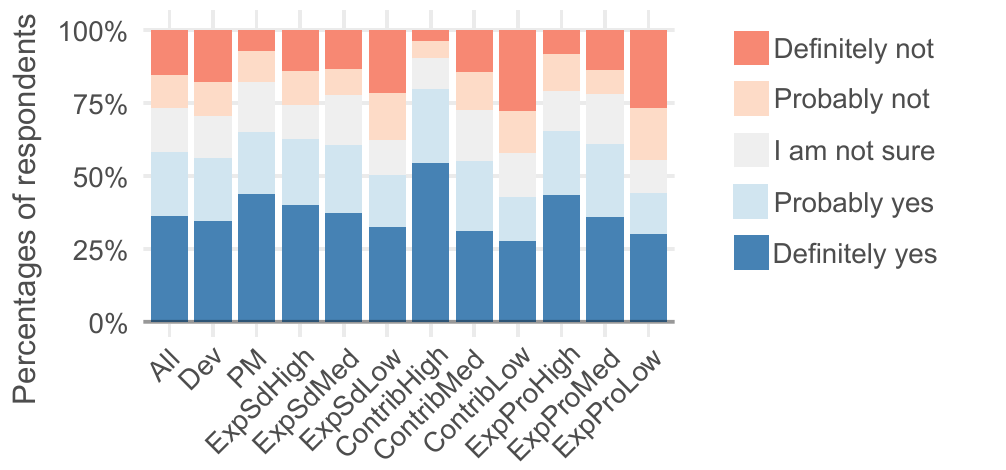}}
\caption{The percentages of self-fixing certainty levels for participants in each demographic group}
\label{fig:rq3_type}
\end{figure}

Fig.~\ref{fig:rq3_type} depicts the certainty of respondents in each demographic group about having self-fixed all five TD types (combined); this is in contrast to Fig. ~\ref{fig:rq3_frequency} which reflects frequency, independently of TD types. For example, in the group \dg{All}, out of a total of 905 ratings, 36.5\%, 21.9\%, 15.0\%, 11.4\%, and 15.2\% respectively were \textit{definitely yes}, \textit{probably yes}, \textit{I am not sure}, \textit{probably not}, and \textit{definitely yes}. As shown in the figure, more than half of the participants in most groups mentioned that they have self-fixed TD (43\%--80\%). However, these percentages are slightly lower than the percentages of the participants who mentioned that they have self-fixed TD \textit{always}, \textit{on a regular basis}, or \textit{sometimes} (i.e., 79\%--97\% in Fig.~\ref{fig:rq3_frequency}). This discrepancy is due to the fact that we focused on individual types of TD when asking about the certainty level; thus, the participants might only self-fix specific types. 

To investigate whether one demographic group tends to self-fix TD more or less frequently than other groups, we conducted pairwise Fisher's exact tests~\cite{fisher_1922} with Bonferroni-Holm corrections~\cite{aickin_1996_adjusting} on the proportions of respondents from the different demographic groups. In detail, those statistical tests were conducted within four sets of demographic groups, i.e., a set of different roles (\dg{Dev} versus \dg{PM}), a set of different levels of software development experience (\dg{ExpSdHigh} versus \dg{ExpSdMed} versus \dg{ExpSdLow}), a set of different contribution levels (\dg{ContribHigh} versus \dg{ContribMed} versus \dg{ContribLow}) and a set of different levels of project experience (\dg{ExpProHigh} versus \dg{ExpProMed} versus \dg{ExpProLow}). Moreover, for the demographic groups in each group set, we conducted eight times pairwise Fisher's exact tests with Bonferroni-Holm corrections to validate the significance of rating differences among these groups. They are the tests on the proportions of different frequencies ratings, together with the attitudes towards self-fixing TD (i.e., \textit{always}, \textit{on a regular basis}, and \textit{sometimes, when it is absolutely essential}) and the negative ones (i.e., \textit{almost never} and \textit{never}). Furthermore, similar to \rqn{1}, we also conducted the tests twice: one to compare \textit{definitely yes} and \textit{probably yes} between the demographic groups; and another to compare \textit{definitely not} and \textit{probably not}. The results per set, are as follows:

\subsubsection{Roles}


We found that 85\% of the participants in the \dg{Dev} group and 96\% of those in the \dg{PM} group demonstrate a rather proactive stance towards self-fixing TD, i.e., they repay their own TD at least when necessary. The result of a Fisher's exact test shows that the difference observed between the two groups is significant ($p$-value = 0.01). Although a similar disparity can be observed in Fig.~\ref{fig:rq3_type}, a Fisher's exact test shows it is not substantial. Altogether, the findings suggest that \finding{people in both types of roles engage in self-fixing TD, although those in project management and support seem to do it more often}.

\subsubsection{Software development experience}

After applying pairwise Fisher's exact tests with Bonferroni-Holm correction, we found no statistically significant difference between the three groups, i.e., \finding{the level of development experience seems to have no effect on how participants repay their own TD}.

\subsubsection{Contributions to the projects}


For the groups with various levels of contribution, we found that 98\% of the participants with high contribution level (\dg{ContribHigh}) have a positive attitude towards self-fixing TD, and that 90\% of them mentioned that they have \textit{definitely} or \textit{probably} self-fixed TD. These percentages were higher than the proportions for practitioners with medium (\dg{ContribMed}) and low (\dg{ContribLow}) contribution levels (87\%/55\% and 80\%/43\% respectively). Similarly, we conducted Fisher's exact tests for the three possible pairwise comparisons (\dg{ContribHigh} versus \dg{ContribMed}, \dg{ContribHigh} versus \dg{ContribLow}, and \dg{ContribMed} versus \dg{ContribLow}). After applying the Bonferroni-Holm correction, we found that participants with high contribution level (\dg{ContribHigh}) differ significantly from the other two groups (\dg{ContribMed} and \dg{ContribLow}) in all of the comparisons ($p$-values \textless 0.05). 

These findings suggest that \finding{practitioners with high contribution level display a significantly more positive attitude towards self-fixing TD}. In addition, more than two-fifths of the participants with low contribution level (42\%) and about one-third of the participants with medium contribution level (27\%) mentioned that they have not self-fixed TD in the projects. As this difference is significant ($p$-value = 0.04), we identified further evidence that \finding{the level of contribution reflects the self-fixing attitude, both positively and negatively}.

\subsubsection{Project experience}


For groups with different levels of experience in the projects, we found that 93\% of participants with higher experience (\dg{ExpProHigh}) consider fixing their own debt at least when necessary. The percentage for those with medium and lower experience (\dg{ExpProMed} and \dg{ExpProLow}) is 89\% and 79\%, respectively. The results of the Fisher's exact tests with Bonferroni-Holm correction suggest a significant difference between participants with high and low experience ($p$-value = 0.02), and no significance between these two groups and \dg{ExpProMed} (medium experience). This finding is further supported by the differences observed in Fig.~\ref{fig:rq3_type}, which are all statistically significant. 
Altogether, these results suggest that \finding{participants who have been involved in the projects for a longer period also seem to be more keen to repay their own TD.}


\section{Discussion}
\label{sec:Discussion}


\subsection{Interpretation of results}

To the best of our knowledge, this is the first study to inquire practitioners on how they repay their own debt. Moreover, we argue that this work is complementary to previous studies that used source code analysis. Thus, we start our discussion with the results that can be compared more directly.

One important finding is that, although most Java and Python practitioners acknowledge that they have self-fixed TD, Java participants were slightly more certain. This may be due to the fact that, 
among the five debt types, 76\%--84\% of Java participants acknowledged that they have noticed similar debt, while the proportions are 53\%--71\% for Python participants. 
We also found that although all participants are likely to acknowledge that they have self-fixed Code Debt, there are some differences between Java and Python practitioners.
For example, the former are similarly concerned about Test Debt, while Python participants have self-fixed Test Debt the least. 
The results are consistent with the findings obtained from source code level, i.e., Test Debt is commonly self-fixed in Java~\cite{liu_ESE2021}, while Test Debt appears to show the lowest self-fixing rate among the five debt types in Python~\cite{Tan2020}. 
We also found that only 4\% of Java participants mentioned that they have not noticed Test Debt, while the percentage of Python participants is over 26\%. One possible explanation may be related to the population of the study: the percentage of Java participants with higher experience and contribution level is about twice that of Python participants; since repaying test smells is time-consuming~\cite{bavota_2015ESE_test}, it may be more commonly performed by experienced developers.

Regarding the reasons to introduce and self-fix TD, one can notice that non-technical reasons seem to be the most cited ones. Almost three-quarters of the reasons for introducing TD are related to \textit{planning and management}, \textit{lack of knowledge}, \textit{people} and \textit{organizational}, while most reasons to self-fix TD are related to \textit{planning and management}, \textit{people} and \textit{knowledge}. The observations related to introducing TD are consistent with the literature~\cite{rios_2020ESE_causes}. There is no literature to compare the self-fixing results, but we do observe that these reasons are quite similar to those for introducing TD.

Despite the dominance of non-technical reasons, we found that technical reasons were frequently mentioned together with non-technical ones. For example, the most recurrent relations involve reasons to introduce TD related to \textit{methodology} and \textit{planning and management}. The latter (non-technical) reason is also the most often associated with reasons to self-fix TD related to \textit{development}.
This intertwining between technical and non-technical reasons reflects the multi-faceted decision-making process related to repaying one's own debt.  
Moreover, we found that practitioners might introduce TD because of \textit{debt inheritance}, which is consistent with a related finding that developers are frequently forced to introduce new TD due to already existing TD~\cite{besker_2018TechDebt} (``the poor are getting poorer''). 

Furthermore, the most cited reason to self-fix TD is the practitioners' \textit{sense of responsibility}, which indicates the relevance of further investigating the impact of human factors on self-fixing TD. The other two most cited reasons for self-fixing TD are to \textit{avoid higher cost} and \textit{improve maintainability}, which indicate a consideration of the trade-off between cost and benefit in order to find the appropriate moment to self-fix TD. This result aligns with related findings that mentioned \textit{focus on short-term goals} and \textit{cost} as reasons when participants are considering whether or not to repay TD ~\cite{freire_2020EASE}.

Finally, the results also suggest that the practitioners' level of involvement (i.e., experience in the project and contribution level) can be related to how they deal with their own TD. 
This observation may be at least partially explained by a sense of code or project ownership: those with higher levels of involvement were significantly more positive towards self-fixing. Furthermore, practitioners in a management role were also found to display a more positive attitude. To assess the impact of practitioners' responsibility on repaying their own TD, we divided the participants into two groups based on whether or not they mentioned \textit{sense of responsibility} as a reason to self-fix TD. The results show that 95\% and 86\% of the participants that did and did not mention responsibility, respectively, have self-fixed TD at least when necessary, although this difference is not statistically significant. In addition, all of the participants that mentioned  \textit{sense of responsibility} acknowledged that they have \textit{definitely} or \textit{probably} self-fixed TD, compared to only half of the other participants. We assessed this disparity with a Fisher's exact test and confirmed it to be significant. Thus, the sense of responsibility is an important driver to self-fix TD and should be properly motivated and rewarded.




\subsection{Implications to researchers and practitioners}

The results suggest that Java and Python practitioners have different attitudes towards self-fixing different types of TD. Thus, researchers can develop tools to prioritize TD remediation by assigning different weights to different TD types and giving suggestions to developers. For example, Test Debt and Defect Debt can be prioritized along with Code Debt for Java projects.
In addition, our findings advance the state of the art as they expand the currently-known list of reasons to introduce TD and provide a list of reasons to repay one's own TD. Broader knowledge of such reasons can lead to more informed decisions when designing future studies on TD. For example, the detailed reasons to introduce and self-fix TD can help researchers in devising and evaluating prevention strategies for the most cited reasons for introducing TD and more timely management strategies for the most cited reasons to self-fix TD. One concrete suggestion is to encourage admission of TD (especially incurred under pressure) and explicitly allocate time to assess the current risk of existing TD.

We learned that practitioners are often willing to address their debt, displaying a sense of responsibility and collaboration. Also, it seems that repayment decisions are not made light-heartedly (e.g., due to pressure or at every single chance) but by weighing various factors, such as time, expected quality and confidence of efficacy. Thus, team leaders should encourage their team to admit TD, discuss it and document the corresponding items in their backlog. 
We also hope that practitioners gain a broader view of possible reasons to introduce and self-fix TD and how common they are; this may help them to make more informed decisions when managing TD and improving software quality.
For example, team leaders can strengthen team communication and promote collaboration, which can boost self-fixing as we observed that ``\textit{improving collaboration}'' is a recurrent reason to self-fix TD. Finally, software engineers should be given incentives to fix their own TD as a means of increasing their technical knowledge, which was also observed as a reason to self-fix TD (\textit{experience and knowledge increased}).


\section{Threats to Validity}
\label{sec:ThreatsToValidity}


\textbf{Construct validity} regards the connection between the research questions and the study objects. Although the survey participants are at the center of the study design, the questionnaire responses are the medium through which we capture their perception. Thus, the study's findings depend on how the participants interpret the questions and how we interpret the responses. To mitigate potential bias in interpretation, we provided definitions of the necessary TD-related concepts.
Also, the level of clarity in answers to open-ended questions (e.g., reasons to self-fix TD) confirms to some extent that the participants understood the underlying concepts.

Furthermore, we used log-transformations in Scott-Knott Effect Size Difference (ESD) test to mitigate the skewness of data distribution; as pointed by Tantithamthavorn et al.~\cite{tantithamthavorn_2016TSE_ESD}, different results might be obtained if other transformations were used (e.g., Blom~\cite{mittas_2013TSE} or Box-Cox~\cite{azzeh_2015JSS} transformations). This limitation is to some extent addressed by repeating each test multiple times to assess if the distribution of bias and variance are approximately normally distributed (as per the Central Limit Theorem)~\cite{moore_2018}.

\textbf{External validity} concerns threats to the generalizability of our findings. Given the population of our study, one cannot expect the results to apply to every development team. 
However, we believe that our findings can be extrapolated to a considerable portion of the Java and Python open-source communities, since our study included participants from popular open-source projects in various domains and with diverse team sizes. While Python and Java are the two most popular programming languages, we cannot generalize our results beyond these languages.

\textbf{Reliability} considers the bias from the researchers in data collection or data analysis. To mitigate bias from the interpretation of open-ended responses (i.e., reasons), we applied open coding in two steps to categorize the replies. Specifically, the first two authors manually classified different reasons of introducing and self-fixing TD. To measure the level of agreement between the classifications of independent researchers, we estimated the inter-rater agreement using Krippendorff's alpha~\cite{krippendorff} to be 0.88\footnote{Krippendorff's inspection of the tradeoffs between statistical techniques establishes that it is customary to require $\alpha\geq0.80$~\cite{krippendorff_alpha}.}, which was calculated based on the classification of all reasons. In the conflicting cases, the first and second authors discussed with the third author until they achieved consensus.
Finally, to support the replication and reproduction of this study, we created a replication package\footnote{\urlrp} with all the necessary data and scripts to run the analyses.


\section{Conclusions}
\label{sec:Conclusion}

This paper reports an empirical study investigating the phenomenon of self-fixed technical debt, i.e., when developers repay TD items that they have introduced themselves.
In particular, we surveyed 181 practitioners spread among 17 projects (written in Java or Python) to explore participants' perception of this phenomenon and under what circumstances they self-fix TD.

The results show that most practitioners display a positive attitude towards self-fixing TD, and seem mostly attentive to Code Debt. Moreover, we found that non-technical reasons (e.g., related to planning and management) are the most considered both when introducing and self-fixing TD, and are often mentioned together with technical reasons (e.g., related to methodology).
In particular, many developers mention a sense of responsibility as a factor for self-fixing, and that decisions to repay are not made easily but by balancing costs and benefits, among other factors.

In the future, we plan to investigate how the reasons portrayed by practitioners can be (semi-)automatically identified through software artefacts such as code, documentation and messages in issue trackers.
We also plan to develop a process and associated tooling to incorporate TD self-fixing in TD-related decisions.

\bibliographystyle{IEEEtran}
\bibliography{mybibfile} 
\end{document}